\documentclass[aps,pre,reprint,groupedaddress]{revtex4-2}
\usepackage[tbtags]{mathtools}
\usepackage{amsmath,amsfonts,mathdots,amssymb,yfonts,calc}
\usepackage{graphicx}
\usepackage{subfigure}
\begin{document}


\title{COLLECTIVE MOTION OF MUTUALLY COUPLED THOMAS OSCILLATORS: SPATIALLY SEPARATED SWIRLING MOTION AND EDDY DIFFUSION}


\author{Vinesh Vijayan}
\email[]{513ph3005@nitrkl.ac.in}
\altaffiliation{Department of Physics and Astronomy }
\affiliation{National Institute of technology, Rourkela, India-769008}

\author{Biplab Ganguli}
\email[]{biplabg@nitrkl.ac.in}
\altaffiliation{Department of Physics and Astronomy }
\affiliation{National Institute of technology, Rourkela, India-769008}

\date{\today}

\begin{abstract}
In this letter, we report a numerical study on the collective dynamics of two mutually coupled Thomas oscillators with linear/nonlinear coupling in a dynamic environment. We claim our model calculations can explain the diffusion of interacting particles in a fluid. In an ordinary fluid, frequent momentum transfer between particles keeps the particles in a fluid moving together with correlated time behaviour. The diffusion of interacting particles in a dynamic environment like this is a nonequilibrium phenomenon and is similar to the observed transient chaotic dynamics in the model. The detailed study of the nature of dynamics and synchronization reveals that, for two qualitatively different regimes of system parameters, the coupled system passes through an interval
of transient chaos before it settles into a chaotic or limit cycle attractor. The linear diffusive coupling is equivalent to weak momentum transfer, leading to conventional dynamics and synchronization. The sinusoidal nonlinear coupling, harmonic momentum transfer, produces exceptional dynamical features. The nature of synchronization is complete(directed motion) when the attractor is chaotic or an unstable transient attractor. In contrast, it is either lag, anti-lag, or space lag for a limit cycle. In such situations, the diffusion is due to particles pedalling and eddy/swirling motion on top of translatory motion via transient chaos. Also, the trajectories of the
two particles in the state space resemble a Chiral Phenomenon.
 
\end{abstract}


\maketitle

\section{\label{sc1}INTRODUCTION}
Relatively simple elements in nature can self-organize into complex behaviours unexpectedly. Why and how these phenomena emerge without a central organizing
entity forms the core of complex system study. Complex systems, often containing more than one individual unit, are open systems that interact among themselves and the environment in a linear/nonlinear way to produce new behaviours (emergent behaviours). The process of achieving these is known as self-organization. Many new dynamical features manifest when the system’s parts interact, such as multistability, transient chaos,  etc. Whether natural or artificial, the complex system interacts to extremize specific processes such that it may be trying to transport, dissipate and distribute energy efficiently.\\  
\\  
Rene Thomas proposed a simple 3D flow that produced unusual dynamical behaviour and became a prototype for chaos studies\cite{th099}. As given in Equation(\ref{thomas}), this model represents a particle moving in a force field with frictional damping under the action of an external source of energy. Two essential characteristics of this system need attention. One is the symmetry under the cyclic interchange of $x, y$, and $z$ coordinates, and the other is the sole parameter $b$, which is the damping coefficient. Originally it was developed as a model for studying the role played by a feedback circuit in generating chaotic behaviour\cite{th099}. Theoretical models based on feedback circuits are critical for understanding cell differentiation and regulatory network\cite{thomas-kauf2}. It is also applicable in chemistry as representative autocatalytic reactions\citep{Ram90}, ecology\citep{Den89}, and in evolution\citep{Stuart}. Under suitable conditions, spatio-temporal patterns are observed for many such oscillators with a nonlinear coupling scheme in \citep{vinesh1} while with a linear coupling and nonidentical oscillators in \citep{Vasi}.
 \\

 \begin{equation}
\begin{split}
\frac{dx}{dt} = -bx +siny\\
\frac{dy}{dt} = -by +sinz \\
\frac{dz}{dt} = -bz +sinx
\end{split}
\label{thomas}
\end{equation} 

\begin{figure}[h]
\begin{center}
{\includegraphics[width = 3.5in,height = 2.5in]{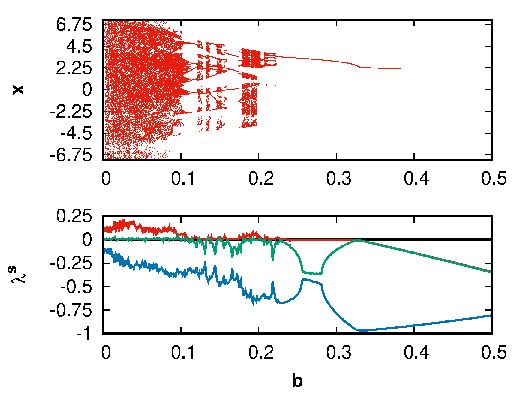}}\\
\end{center}
\caption{Bifurcation and Lyapunov spectrum for Thomas oscillator.}
\label{Fig_thomas}
\end{figure}

\noindent When the generalized coordinates are components of velocities, one can think about the model as a Brownian particle with its immediate fluid environment as a dynamical system. The nature of the force field depends upon the value of $b$, the damping parameter. This forcing term is due to the local fluctuation of velocities which is highly vigorous at low damping values, thereby producing a turbulent environment for the particle. Therefore, the particle’s velocity is unpredictable due to its chaotic nature for low damping coefficient values. The velocity at this instant is not sufficient to predict the velocity in subsequent times. Thus, instantaneous fluctuating velocity components are usually unknown, generating randomness due to chaoticity. By regulating the parameter, one can tune from a highly turbulent force field at low $b$ to a highly ordered force field at higher values $b$. Hence the particle motion at low values of $b$ will be chaotic, and at high values, it is stationary. Figure(\ref{Fig_thomas}) shows the bifurcation and Lyapunov spectrum as a function of $b$. The system varies smoothly from a chaotic dissipative system to a chaotic conservative system. on decreasing $b$. The latter case provides the only example for fractional diffusion in a purely deterministic system\cite{th099}\cite{Vasi}. In the limit $b=0$, the system can perform a chaotic walk. Route to chaos and its symbolic dynamics are well-studied in \cite{spo07}\cite{row08}.
\\
\\
In the present work, we consider two identical Thomas oscillators under a mutual coupling scheme with linear and non-linear coupling functions. Though the majority of the studies consider linear coupling in chaotic systems, the non-linear coupling is general and more applicable to real systems\cite{Ait18}. Linear coupling is an approximate case of non-linear coupling, like in harmonic approximation. The velocity-velocity coupling introduced here stands for momentum transfer. The linear diffusive coupling means weak momentum transfer. While nonlinearly diffusive coupling stands for a  strong momentum transfer between the particles. The motivation is to look at the subsequent collective dynamics and characterize them. For the study of collective motion, we have chosen $b=0.18$ and $b=0.1998$. These values of $b$ generate stable chaotic oscillations with a reasonable attractor, unlike attractors of enormous sizes for shallow values of $b$. Further, the size of the attractor at $b=0.1998$ is just one-third of the size of the attractor at $b=0.18$. Moreover, the uncoupled systems do not show any transient chaos for these system parameter values. There is one more major difference between the systems with $b =0.18$ and $b=0.1998$. The dynamics in the latter case is such that the stable chaotic dynamics fall at the boundary of chaos and quasi-periodic oscillations. In this sense, this value of $b$ is for comparison purposes (Figure(\ref{Fig_thomas})). We call this regime chaotic at the boundary of chaos and quasi-periodic oscillations.
\\
\\

\noindent The coupled system dynamics are explained based on the Largest Lyapunov Exponent(LLE) and bifurcation diagram. The qualitative and quatitative measure of synchronization like LTE, Pearson and TD are well established measures and are discussed in details in \cite{fuji83}\cite{gon04}\cite{gau96}. The directed motion of particles, in this case, can be seen as a complete synchronization process(CS). Similarly, the Pearson coefficient($\rho$) and the transverse distance (TD) on the synchronization manifold probe the velocity-velocity correlation and the onset of Complete Synchronization(CS). Synchronization thresholds are analyzed and compared for the two cases with linear and nonlinear coupling functions.\\
\\
The organization of the paper is in the following way. In (\ref{sc2}), there is a brief discussion on the observed transient chaotic phenomena with the linear and nonlinear coupling of the oscillators.  (\ref{sc3}) contains a discussion on the dynamics and velocity-velocity correlation of two coupled oscillators with linear coupling for different parameter values. In (\ref{sc4}), the same discussion is there for nonlinear coupling under mutual coupling of the oscillators. Finally, in (\ref{sc5}), we present results and related discussions.
\section{\label{sc2}Mutually Coupled Thomas System and Transient Chaos}
\noindent Traditionally chaos is a long-term asymptotic property that lives for, from a physical point of view, time scales much larger than the largest observational time scales. However, chaotic phenomena with a finite life span are known as transient chaos, such that the time scales are shorter than the largest observational time. Such phenomena are interesting and relevant in many applications\cite{Tel}\cite{Didier}\cite{Mccann}\cite{Dhamala}\cite{Ravasz}\cite{Sumi}. Depending upon the initial conditions, these transients live for different times in different systems\cite{Tel}\cite{Tel1}\cite{Ruben}. These transient states are examples of a non-equilibrium state which is different from the asymptotic state. Wherever transient chaos is present, the system moves around chaotically and then suddenly jumps to a steady-state, different from the transients.\\ 
\\
\begin{figure}[hbt!]
    \centering
    \subfigure[]{\includegraphics[height=4.7cm,width=6.2cm]{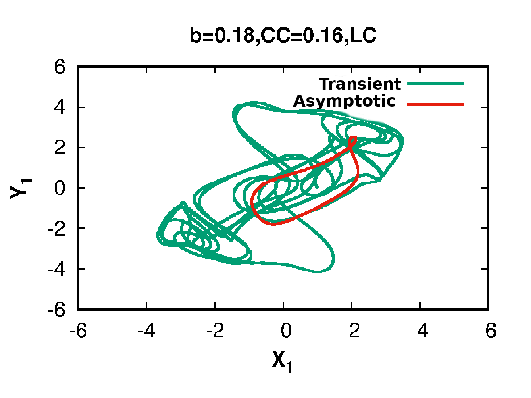}} \\    
    \subfigure[]{\includegraphics[height=4.7cm,width=6.2cm]{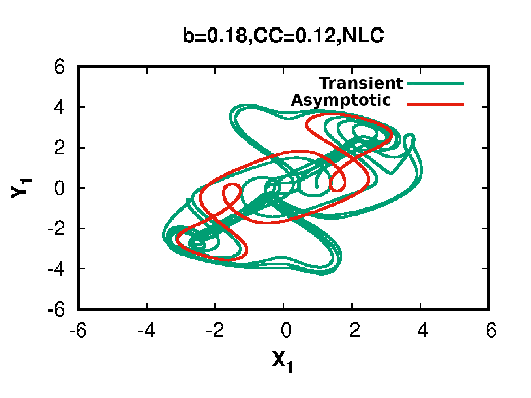}}     
  
      \caption{Plots showing transient chaos (green) and stable at-
tractor(red) for linear(LC) and nonlinear(NLC) coupling for the choice of parameter values 0.18 and 0.1998 with specified coupling coefficients.} 
 \label{CH6F22}
\end{figure}

\noindent Two Thomas oscillators, coupled either linearly or non-linearly(sinusoidal),  undergo a long period of transient, unstable attractor, mainly chaotic, before settling to a stable attractor. The transient period depends on the initial condition. The transient period will get reduced with a suitable initial condition, but it remains reasonably long. The period of transient chaos is longer in the case of $b = 0.1998$ than $b = 0.18$. The period gets further increased due to non-linear coupling.\\
\\
\noindent We use three different initial conditions to study transient states and explore possible final asymptotic states. (i) The first one is what we call resetting. In this case, a fixed identical initial condition is there for all calculations with different coupling constants. The initial condition is chosen from the basin of attraction of the uncoupled oscillators. This case is equivalent to performing independent experiments on the same system but with a different coupling constant fixing an identical initial state. (ii) In the second case,  we start with the same initial condition as in the first case for an uncoupled system. For the subsequent coupled system, the long-term solution of the previous coupled system is varied by $10\%$ to be chosen as the  initial condition for the next coupled system with a different coupling constant. This is done by varying the coupling constant by a step of $ \delta cc$ to achieve a new coupled system. (iii) In the third case, $1\%$ variation is applied to the previous long-term solution for the next system. It is essential to see how to control these transients for practical purposes. The last two cases require a single experiment with a variable coupling element that will vary during an experiment, like a variable resistor in a circuit. When the coupling element is varied simultaneously, one of the oscillators is slightly disturbed externally. The initial condition for the new coupling is $10\%$  or $1\%$ different from the final state of the previous run.\\
\\
\noindent (Figure(\ref{CH6F22})) shows the transient chaos observed at the periodic(limit cycle) windows with the third set of initial conditions. For $b=0.18$ with linear coupling and at $cc=0.16$, we find the asymptotic state to be a limit cycle with transient time $5K$ of chaos(Figure(\ref{CH6F22}-a)). With non-linear coupling, at $cc=0.12$, we observe transient chaos, and the asymptotic state is again a limit cycle corresponding to complex oscillations but with transient time $25K$. So, non-linear coupling enhances the transient period by five times, much longer(Figure(\ref{CH6F22}-b)).\\
\\ 
One can prove the existence of transient chaos in the coupled system by simply plotting the power spectrum below and above the transient time. There should be a spread in the frequency distribution for transient chaotic regimes, while in the case of regular trajectories like a limit cycle, there will be distinct spikes at harmonics. 
\begin{figure}[hbt!]
    \centering
    \subfigure[]{\includegraphics[height=4.5cm,width=4.2cm]{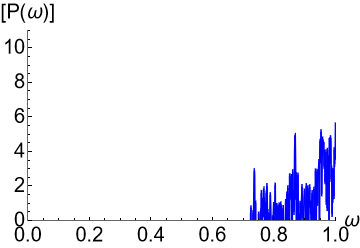}}     
    \subfigure[]{\includegraphics[height=4.5cm,width=4.2cm]{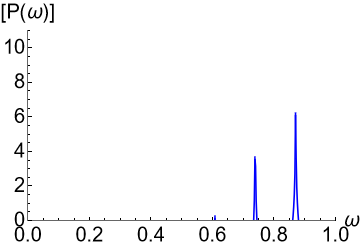}}    
    
      \caption{Power spectrum for  b=0.18 with linear coupling and cc = 0.16(a)  Transient case (b) Asymptotic case.}
\label{CH6FPWS}
 
\end{figure}
Figure(\ref{CH6FPWS}) shows the power spectrum for the case discussed in (Figure(\ref{CH6F22}-a)), and it shows that the asymptotic case is a period two limit cycle. Similarly, one can also distinguish between the asymptotic and transient cases for the remaining cases.\\
\\
\section{\label{sc3}Mutually Coupled Thomas System With Linear Coupling}
The mutual linear coupling is provided to the $x$ variable. Since the system is symmetric to interchange among all the variables, coupling to other variables instead of $x$ would give the same dynamics. The governing equation of motion is given by 
\begin{equation}
\begin{split}
\dot{ x}_{1,2} &=-bx_{1,2} + \sin y_{1,2} +cc*(x_{2,1}-x_{1,2})\\
\dot{ y}_{1,2} &=-by_{1,2} + \sin z_{1,2}\\
\dot{ z}_{1,2} &=-bz_{1,2} + \sin x_{1,2}
\end{split}
\label{CH6E1}
\end{equation}
where $cc$ is the coupling coefficient.  In this case, the final state does not depend on the initial condition showing the absence of multistable states. We choose the third set of initial conditions for all the calculations for this choice, and transient time is the least. 
\subsection{Linear diffusive coupling with b=0.18}
\begin{figure}[hbt!]
    \centering
    \subfigure[]{\includegraphics[height=3.5cm,width=8.0cm]{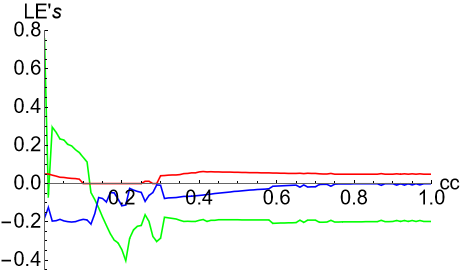}} \\    
    \subfigure[]{\includegraphics[height=3.5cm,width=8.0cm]{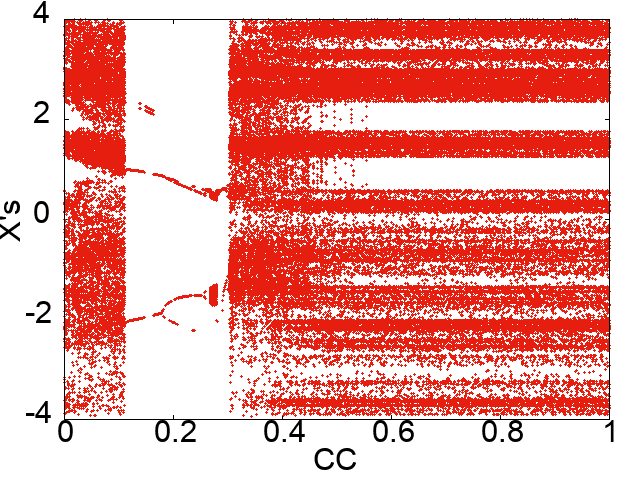}}     
   
      \caption{Lyapunov spectra and bifurcation diagram for the mutually coupled Thomas system with linear diffusive coupling with $b=0.18$(stable chaotic oscillations).} 
 \label{CH6F1}
\end{figure}

The dynamics of the mutually coupled system with $b=0.18$ can be visualized from the Lyapunov spectra( three largest) and bifurcation diagram, as shown in the Figure(\ref{CH6F1}). The nature of the dynamics is such that there is a transition from hyper-chaotic to chaotic behaviour, indicated by the number of positive Lyapunov exponents, as we smoothly vary $cc$ from zero to a sufficiently large value. The nature of dynamics also changes for a range of intermediate values of the coupling coefficient. There are limit cycle oscillations, as indicated by the zero value of LLE(red line ), with all others being negative. In the range  $cc \sim 0.11$ to $cc \sim 0.3$, there is a small window of chaos confirmed by the positivity of the LLE(red line).
\\

\noindent The negativity of LTE gives the stability of the synchronized manifold. LTE is calculated from the time evolution of the perturbation projections perpendicular to the synchronized manifold. Parallel and perpendicular projections of  perturbations are defined by Equation(\ref{perturbation}) and their time evolutions are governed by Equation(\ref{stability_linear}).
 \\
\begin{equation}
{\delta {\bf X_{\parallel}} = (\delta {\bf x_1} + \delta {\bf x_2})/\sqrt{2}\hspace{0.8cm}
\delta {\bf X_{\perp}} = (\delta {\bf x_1} - \delta {\bf x_2})/\sqrt{2}}
\label{perturbation}
\end{equation}

\begin{equation}
{\delta \dot{\bf X}_{\parallel} = {\bf J(X)}.\delta {\bf  X}_{\parallel}\hspace{1.0cm}
\delta \dot{\bf X}_{\perp} = {[\bf J(X)- 2C]}.\delta {\bf  X}_{\perp}}
\label{stability_linear}
\end{equation}
where {\bf J(X)} is the Jacobian matrix evaluated at the synchronization manifold. ${\bf C}$ is the coupling matrix given by
\begin{equation}
{\bf C} = cc \begin{bmatrix}
1 & 0 & 0\\
0 & 0 & 0\\
0 & 0 & 0\\
\end{bmatrix}
\label{CH6E4}
\end{equation} 
where $cc$ is the coupling strength.
\noindent The synchronization manifold is stable when the perturbations transversal to the manifold die out exponentially. This implies that all the Lyapunov exponents for the transversal perturbations are negative\cite{fuji83}.\\

\begin{figure}[hbt!]
\begin{center}
{\includegraphics[width = 3.2in,height = 1.5in]{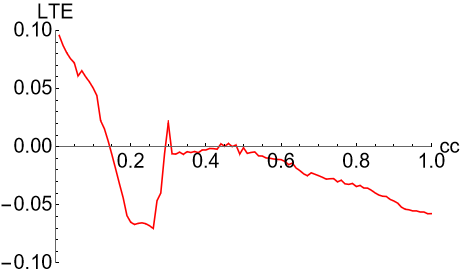}}
\end{center}
\caption{ Transverse Lyapunov exponent $\lambda{^\bot}$(LTE) for the mutually coupled Thomas system with linear diffusive coupling with $b=0.18$(stable chaotic oscillations).}
\label{CH6F2}
\end{figure}

\noindent The system displays a transition to complete synchronized(CS, directed motion) state, defined by $\mathbf{ x_1}(t) = \mathbf{ x_2}(t) =\mathbf{X}(t)$, at a critical value of $cc_T$. CS is the only type of synchronized state-observed throughout the range of $cc$ from the onset(critical) value. This transition is due to two counter-balancing effects. The LTE measures the instability of a synchronized manifold, whereas the diffusion measures the stability. When diffusion dominates over instability, the system synchronizes, and the motion takes place in an invariant subspace of synchronization manifold described by $ \mathbf{\dot{X}}(t)= \mathbf{F}(\mathbf{X}(t))$. Detection of CS is observed  by the calculation of Pearson coefficient($\rho$), the degree of cross-correlation between the variables\cite{gon04}, defined by Equation(\ref{pearson_linear}).\\
\begin{equation}
{  \rho =\frac{\langle(x_1-\langle x_1 \rangle)(x_2 -\langle x_2 \rangle)\rangle}{\sqrt{\langle(x_1-\langle x_1 \rangle)^2\rangle \langle(x_2 -\langle x_2 \rangle)^2 \rangle}}}
\label{pearson_linear}
\end{equation}
where $<.>$ denotes full space time average. The averages are calculated after the initial transients. $\rho =1$  shows that the two variables are completely correlated, and $\rho = -1$ is negatively correlated. $\rho = 0$ indicates that the two variables are completely uncorrelated. We rely on another important measure of the CS manifold, that is, the average distance from the synchronized manifold $|x_\perp|_{rms}$ and the maximum observed values $|x_\perp|_{max}$. The former is sensitive to global stability while the latter to local stability\cite{gau96}. It is defined by Equation(\ref{average_dis}).

\begin{equation}
{  |x_\perp|_{rms} =\lim_{T \to \infty} \frac{1}{T -T_0} \int_{T_0}^T| x_1(t) -x_2(t)| dt}
\label{average_dis}
\end{equation}
where $T_0$ is the transient time and T, is the total time of computation.\\
\begin{figure}[hbt!]
\begin{center}
{\includegraphics[width = 3.5in,height = 2.2in]{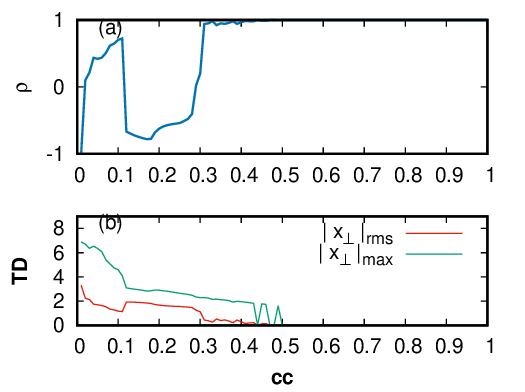}}
\end{center}
\caption{ Pearson coefficient ($\rho$) and Transverse distance (TD) plot for the mutually coupled Thomas system with linear diffusive coupling with $b=0.18$(stable chaotic oscillations).}
\label{CH6F3}
\end{figure}

\noindent From Figure(\ref{CH6F3}), the threshold for CS is confirmed to be $cc_T \sim 0.49$ which is stable as indicated by the plot for TLE, Figure(\ref{CH6F2}). From Figure(\ref{CH6F2}), we also make the following observations that TLE is negative in the range $cc \sim 0.14$ to $cc \sim 0.3$ and $cc \sim 0.31$ to $cc \sim 0.43$ such that the former is well within the limit cycle region as observed from the Lyapunov spectra.  They both correspond to weak forms of stable synchronization, which is indeed possible on the way to achieving CS. Figure(\ref{CH6F3}) shows Pearson coefficient($\rho$) and Transverse distance (TD) and agrees well with TLE.\\
\\
\noindent The above picture of dynamics and synchronization is not affected by changing the initial condition. The only change due to different initial conditions is the survival time of transient chaos. In the present calculation, transient chaos persisted up to $T=5K$ with the $1 \%$ variation in the initial condition after every $\delta cc$ cycle. The critical point to be noted here is that the transient states do not affect the synchronization. The synchronized and desynchronized states are shown below in Figure(\ref{CH6F4}).

\begin{figure}[hbt!]
\begin{center}
{\includegraphics[width = 3.5in,height = 2.8in]{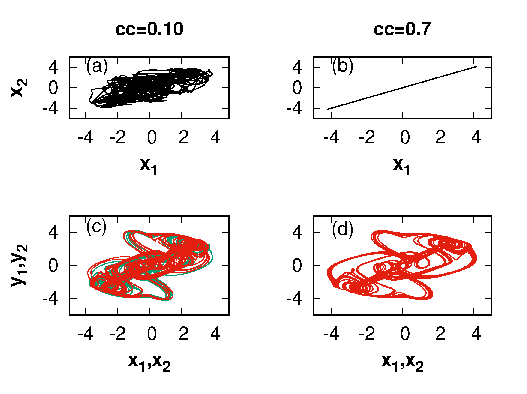}}
\end{center}
\caption{ Phase portrait in the case of Linear Coupling(LC) with $b=0.18$ for two different coupling coefficient values. Desynchronized state (left panel) and synchronized state(right panel).
}
\label{CH6F4}
\end{figure}
\subsection{Linear diffusive coupling with b=0.1998}
The qualitative nature of the dynamics remains the same as in the previous case. The small chaos window is now in the middle of the limit cycle region in the range of $cc \sim 0.12$ to $cc\sim 0.29$. For synchronization, the threshold for CS  shifted towards the left with a value of $cc_T \sim 0.35$. Stable CS is achieved much earlier than in the previous case and confirmed by the LTE. Like the previous case, we also have weak forms of synchronization in the range of limit cycle oscillations in the low range of $cc$. In the present calculation, transient chaos persisted up to $T=50K$ with $1 \%$ variation in the initial condition after every $\delta cc$ cycle. Here also, synchronization does not get affected by the transient states.  \\ 
\\
So for both these parameter values, linear coupling(weak momentum transfer) results in directed motion of the particles for higher values of the coupling constants. The external energy from thermal fluctuation and local kinetic fluctuations are converted into translatory motion of the two particles. This is quite similar to molecular diffusion. The asymptotic dynamics of the coupled system remain chaotic for strong coupling values.
\section{\label{sc4}Mutually Coupled Thomas System With Nonlinear Coupling}
In the case of sinusoidal nonlinear bidirectional coupling, the dynamics is given by 
\begin{equation}
\begin{split}
\dot{ x}_{1,2} &=-bx_{1,2} + \sin y_{1,2} +cc*\sin(x_{2,1}-x_{1,2})\\
\dot{ y}_{1,2} &=-by_{1,2} + \sin z_{1,2}\\
\dot{ z}_{1,2} &=-bz_{1,2} + \sin x_{1,2}
\end{split}
\label{CH6E2}
\end{equation}
The dynamics of transverse perturbation, governed by the Equation(\ref{stability_linear}) gets modified to (without any approximation (linearization))
\begin{equation}
\begin{split}
\delta \dot{\mathbf X}_{\perp} = {[ J(\mathbf{X})- 2*{\bf C}*\cos({x_2 - x_1})]}.\delta {\mathbf  X}_{\perp}
\end{split}
\label{CH6E3}
\end{equation}
\subsubsection{Nonlinear diffusive coupling with b = 0.18}
\begin{figure}[hbt!]
    \centering
    \subfigure[]{\includegraphics[height=3.5cm,width=8.0cm]{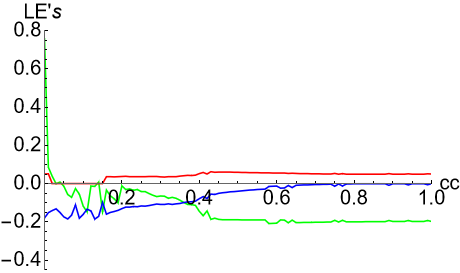}} \\    
    \subfigure[]{\includegraphics[height=3.5cm,width=8.0cm]{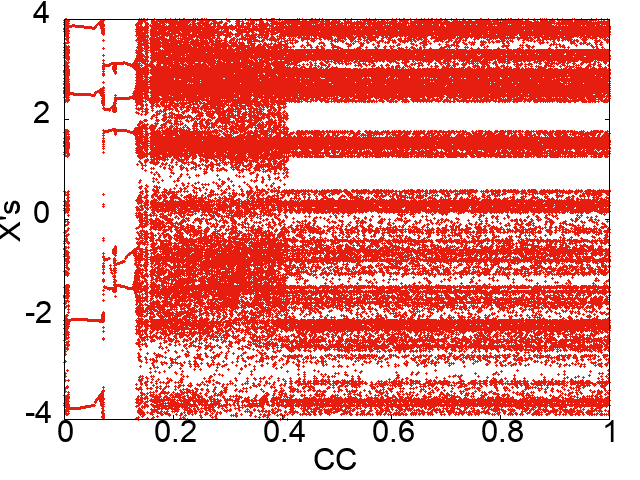}}     
   
      \caption{Lyapunov spectra and bifurcation diagram for the mutually coupled Thomas system with nonlinear diffusive coupling with $b=0.18$(stable chaotic oscillations).} 
 \label{CH6F9}
\end{figure}

In this case, one can immediately see the effect of nonlinear coupling compared to the linearly coupled case. The range of $cc$ of the limit cycle oscillations now shifted to the left, $0.012-0.13$. The dynamics are such that there is a transition from hyper-chaotic to chaotic behaviour via limit cycle oscillations. Also, the small region of chaotic oscillations in the limit cycle region has disappeared. The Lyapunov spectra and bifurcation diagrams are  in  Figure(\ref{CH6F9}). The third set of initial conditions achieves this with a transient time of $25K$.\\

\begin{figure}[hbt!]
\begin{center}
{\includegraphics[width = 3.2in,height = 1.5in]{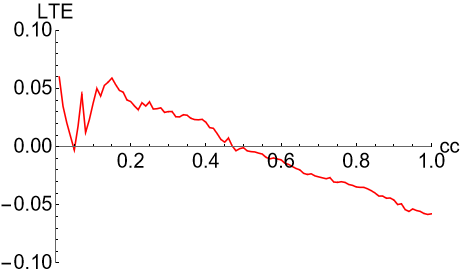}}
\end{center}
\caption{ Largest Transverse Lyapunov exponent $\lambda{^\bot}$(LTE) for the mutually coupled Thomas system with nonlinear diffusive coupling with $b=0.18$(stable chaotic oscillations).}
\label{CH6F10}
\end{figure}

\begin{figure}[hbt!]
\begin{center}
{\includegraphics[width = 3.5in,height = 2.5in]{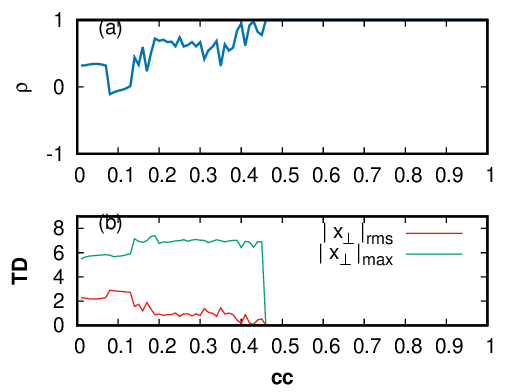}}
\end{center}
\caption{ Pearson coefficient ($\rho$) and Transverse distance (TD) plot for the mutually coupled Thomas system with nonlinear diffusive coupling with $b=0.18$(chaotic oscillations).}
\label{CH6F11}
\end{figure}

\noindent The threshold for  CS, in this case, is found to be $cc_T = 0.46$, which is slightly less than that of the case with the linear coupling, and it is stable, as indicated by LTE in Figure(\ref{CH6F10}). Figure(\ref{CH6F11}) shows the threshold value for synchronization. Before this transition,  LTE becomes negative at a particular value of $cc = 0.05$, where LLE simultaneously touches the zero value, showing a weak form of synchronization, which is amplitude envelope synchronization(AES)\cite{gon02}. However, AES is extremely sensitive to the changes in coupling constant, as evident from the LTE spectrum.\\
\\

\begin{figure}[hbt!]
\begin{center}
{\includegraphics[height=3.5cm,width=8.0cm]{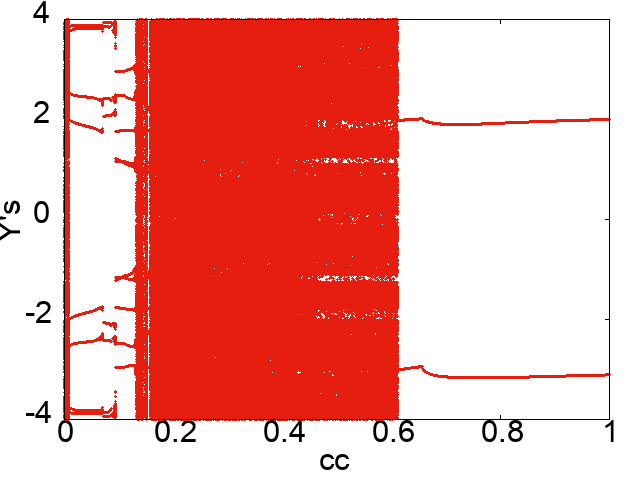}}
\end{center}
\caption{ Bifurcation diagram for $1\%$ variation in the initial condition after every cycle of $cc$ with nonlinear diffusive coupling with $b=0.18$ and transient time 24k(chaotic oscillations).}
\label{CH6F12}
\end{figure}

\begin{figure}[hbt!]
\begin{center}
{\includegraphics[width = 3.5in,height = 2.5in]{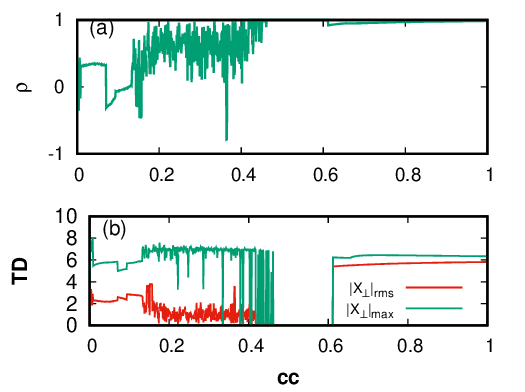}}
\end{center}
\caption{Pearson coefficient ($\rho$) and Transverse distance (TD) plot for the mutually coupled Thomas system with nonlinear diffusive coupling with $b=0.18$ and transient time 24k (chaotic oscillations).}
\label{CH6F13A}
\end{figure}

\noindent Calculations with the second set of initial conditions for the nonlinear coupling show a transient period of $25K$. However, we get the same picture of asymptotic dynamics and the state as chaotic. Interesting transient phenomena are observed when we look at the dynamics below the transient time, say $24K$, but with the third set of initial conditions, shown in Figure(\ref{CH6F12}). Here for lower values of $cc$, the dynamics are almost the same as that of the $25K$ transient case. However, at higher values of coupling coefficients, we have different dynamical behaviour. Now beyond $cc=0.61$, we have limit cycles. Such behaviour is the signature of non-equilibrium phenomena in diffuion of particles in a fluid. The synchronization property calculated for this case with the Pearson coefficient and transverse distance shows an exciting result, as shown in Figure(\ref{CH6F13A}).\\
\\

\begin{figure}[hbt!]
\begin{center}
{\includegraphics[width = 3.0in,height = 1.8in]{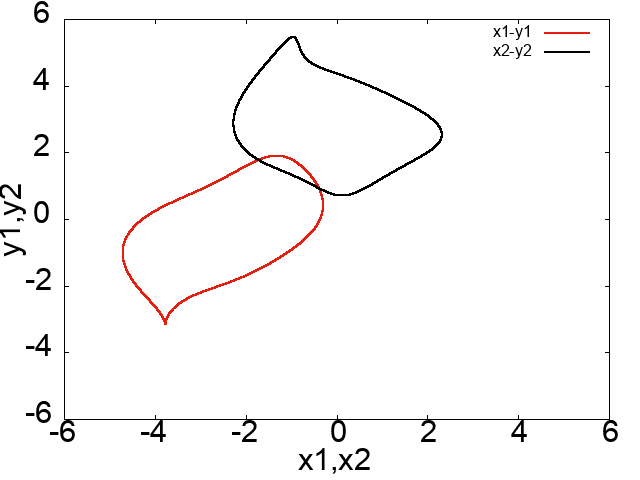}}
\end{center}
\caption{The projected state space of particle one and two at $cc=0.7$ for the mutually coupled Thomas system with nonlinear diffusive coupling with $b=0.18$ and  transient time between 24K and 25K (chaotic oscillations).}
\label{CH6F13BC}
\end{figure}

\noindent 
From $\rho$ and TD calculations, we note the following observations. There is CS in the range of $cc = 0.46$ to $cc= 0.61$. After that, there is no CS, but a strong correlation exists between the variables as $\rho \approx 1$. Time series analysis for the range $cc=0.61$ to $cc=1$ shows the oscillations in this range are spatially separated, or we can say there is a space lag. We observe two independent nearly loopy motions(limit cycle) of the velocities with different orientations in the state pace. The speeds are closely correlated, as shown by the Pearson coefficient, and it is not CS as indicated by the TD plot. We call this a space lag synchronization or eddy/swirling motion and manifests turbulent flow in the environment, and the diffusion process is known as eddy diffusion. The projected state space for particles one and two are shown in Figure(\ref{CH6F13BC}). The two trajectories in the state space are not exactly mirror images of each other and resemble a ”Chiral phenomenon”. One cannot take the trajectories and superimpose them by a rotation and translation, and so there is no CS for these cases. 

\noindent This is indeed possible in the case of diffusion of coupled particles in a dynamic environment, and it shows the richness of the transient phenomena. A similar result is found with the second set of initial conditions too.

\subsubsection{Nonlinear diffusive coupling with b = 0.1998}
\begin{figure}[hbt!]
    \centering
    \subfigure[]{\includegraphics[height=3.5cm,width=8.0cm]{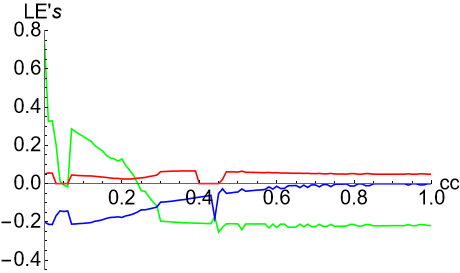}} \\    
    \subfigure[]{\includegraphics[height=3.5cm,width=8.0cm]{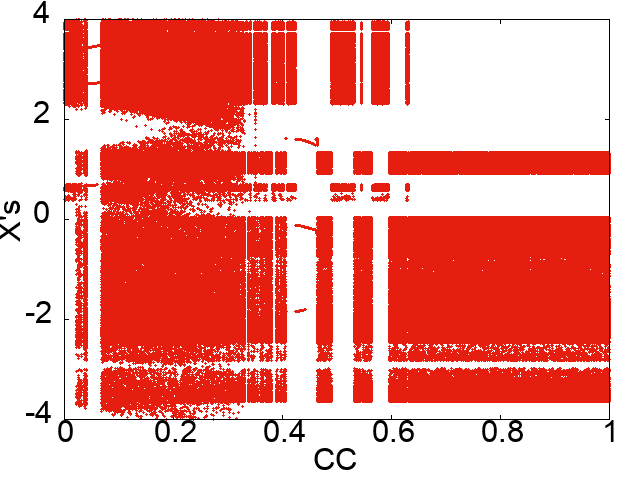}}     
   
      \caption{Lyapunov spectra and bifurcation diagram for the mutually coupled Thomas system with nonlinear diffusive coupling with $b=0.1998$(stable chaotic but at the border of chaos and quasi-periodic oscillations).} 
 \label{CH6F13}
\end{figure}

Finally, we look at the synchronization properties of mutually coupled Thomas systems under sinusoidal coupling for $b = 0.1998$ with the third set of the initial condition. The dynamics were found to be different from the previous cases. Now we have two windows of limit cycle oscillations, i.e. the dynamics become much more complex in this case. The two windows of limit cycle oscillations are $cc \sim 0.029$ to $cc \sim 0.059 $ and $cc \sim 0.42$ to $cc \sim 0.46 $. The final state, achieved by the coupled system,  is chaotic but after a much longer transient time, which is $80K$. The hyper-chaotic nature of the coupled system persists much longer here, up to $cc=0.23$ except for the first window of the limit cycle oscillation. The bifurcation diagram and Lyapunov spectra are shown in Figure(\ref{CH6F13}).\\

\begin{figure}[hbt!]
\begin{center}
{\includegraphics[width = 3.2in,height = 1.5in]{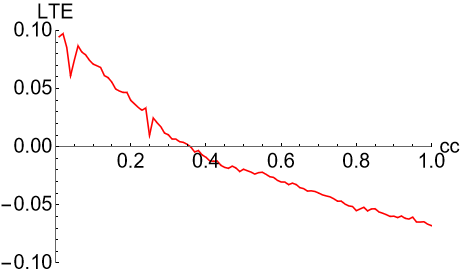}}
\end{center}
\caption{ Largest Transverse Lyapunov exponent $\lambda{^\bot}$(LTE) for the mutually coupled Thomas system with nonlinear diffusive coupling with $b=0.1998$(stable chaotic but at the border of chaos and quasi-periodic oscillations).}
\label{CH6F14}
\end{figure}

\begin{figure}[hbt!]
\begin{center}
{\includegraphics[width = 3.5in,height = 2.2in]{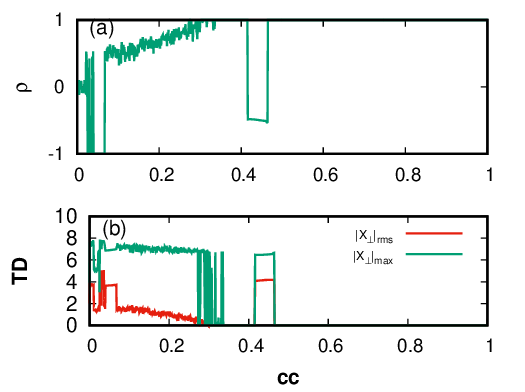}}
\end{center}
\caption{ Pearson coefficient ($\rho$) and Transverse distance (TD) plot for the mutually coupled Thomas system with nonlinear diffusive coupling with $b=0.1998$(stable chaotic but at the border of chaos and quasi-periodic oscillations).}
\label{CH6F151}
\end{figure}

\begin{figure}[hbt!]
\begin{center}
{\includegraphics[width = 3.5in,height = 3.0in]{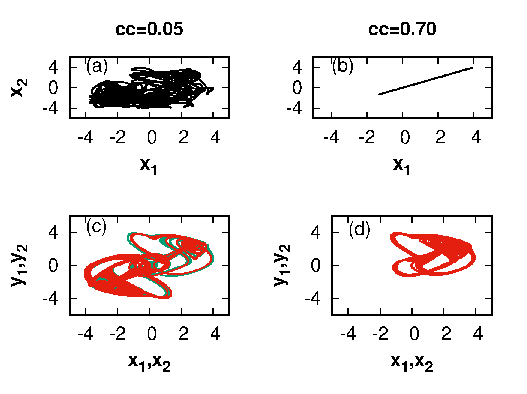}}
\end{center}
\caption{Phase portrait in the case of nonlinear Coupling(NLC) with $b=0.1998$ for two different values of coupling coefficient. Desynchronized state(left panel) and synchronized state(right panel).}
\label{CH6F16}
\end{figure}

\noindent The CS threshold is around $cc_T \sim 0.35$ and is stable, as indicated by the LTE in Figure(\ref{CH6F14}). In this case, there is only a single cut along the coupling axis on the way to stability. The Pearson coefficient and Transverse distances for this case are shown in Figure(\ref{CH6F151}). The phase portraits show the desynchronized state as well as the synchronized states for two different values of coupling coefficient, as shown in Figure(\ref{CH6F16}).
\\

\begin{figure}[hbt!]
    \centering
    \subfigure[]{\includegraphics[height=3.5cm,width=6.5cm]{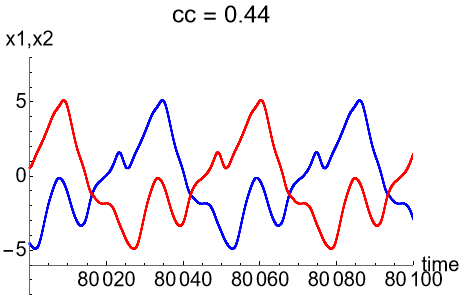}} \\    
    \subfigure[]{\includegraphics[height=3.5cm,width=6.5cm]{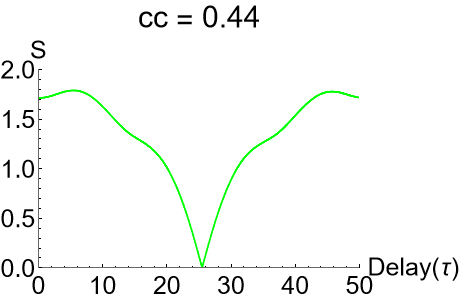}} \\   
    \subfigure[]{\includegraphics[height=3.5cm,width=6.5cm]{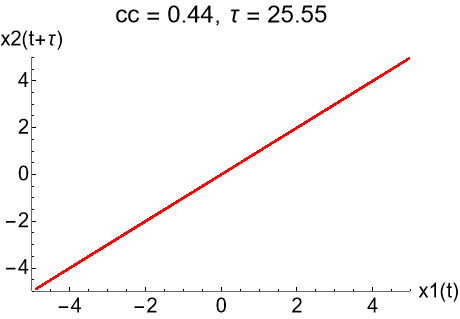}}     
    
  \caption{(a)The time series,(b)Similarity index and (c)Lag synchronization for nonlinear coupling with $b=0.1998$ and $cc=0.44$} 
 \label{CH6F19}
\end{figure}

\noindent  In this case, we do not observe complete synchronization in the second limit cycle range, that is, $cc=0.42$ to $cc=0.46$ . To predict the nature of synchronization, we choose a specific value in this range, say, $cc= 0.44$, and we analyze the time series. Figures(\ref{CH6F19}) show there are indeed co-operative dynamics but not CS at these values of coupling. The synchronized states for $cc=0.44$  seem to be Lag Synchronization(LS). To quantify and visualize the LS/ALS, we use the idea of similarity function\cite{ros97}. The similarity function is defined as the time average of the variables $x_1(t)$ and $x_2(t+\tau)$. It is given by Equation(\ref{similar}). The LS can be seen by plotting $x_2(t+\tau)\hspace{.1cm}vs\hspace{.1cm}x_1(t)$, Figure(\ref{CH6F19}), and ALS by plotting $x_2(t-\tau)\hspace{.1cm}vs\hspace{.1cm} x_1(t)$. We call this a "pedalling motion". The pedalling motion is the two nearly circular motions( limit cycles )with a constant time shift(phase lag) and a common axis of revolution. The velocity of particle one is followed by the other with a phase difference. In eddy motion, there are two independent axes of revolution. Pedalling motion manifests a complex diffusion process as eddy motion. In the second limit cycle range, one can observe either LS or ALS.
 \\
 \begin{equation}
\begin{split}
S^2_\mp =\frac{\langle{(x_2(t+\tau)} \mp { x_1(t)})^2 \rangle}{\sqrt{\langle{x_1}^2(t)\rangle \langle{x_2}^2(t)\rangle}}
\end{split}
\label{similar}
\end{equation}

\begin{figure}[hbt!]
\begin{center}
{\includegraphics[height=3.5cm,width=8.0cm]{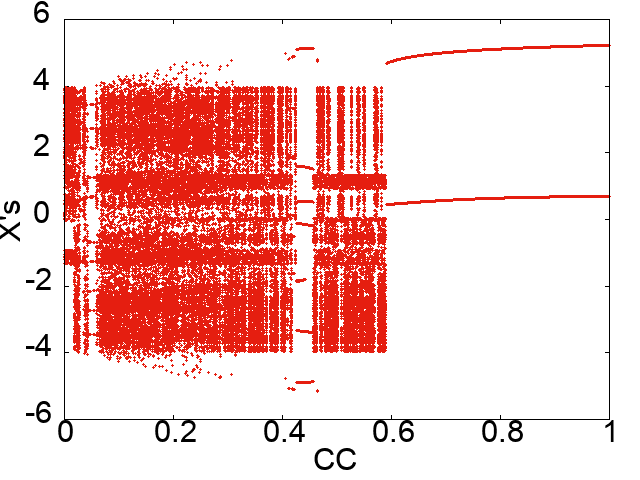}}
\end{center}
\caption{ Bifurcation diagram for the mutually coupled Thomas system with nonlinear diffusive coupling with $b=0.1998$ (chaotic oscillations at the boundary of chaos and quasi-periodic) with $10\%$ variation in initial condition after every cycle of $cc$, Transient time 80k.}
\label{bfx_1998NL_10}
\end{figure}

\begin{figure}[hbt!]
\begin{center}
{\includegraphics[width = 3in,height = 2.5in, angle = 0]{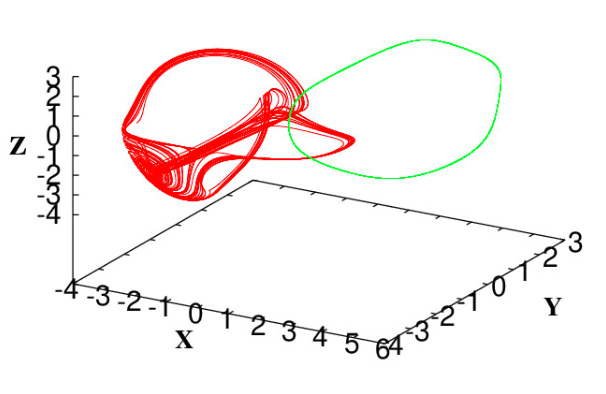}}
\end{center}
\caption{ Phase space plot for different initial conditions: with $10\%$ variation (limit cycle) and $1\%$ variation (chaotic) in the initial condition after every cycle of $cc$ with nonlinear diffusive coupling with $b=0.1998$, Transient time 80k(chaotic oscillations at the boundary of chaos and quasi-periodic).}

\label{multi}
\end{figure}

\noindent Our calculation with the second choice of initial condition shows that  synchronized state after the critical value of $cc = 0.59$ is  a stable limit cycle instead of chaos, shown in the bifurcation diagram(Figure(\ref{bfx_1998NL_10})). The limit cycle occupies largely different regions in the phase space than the transient chaotic attractor, as shown in Figure(\ref{multi}) and shows the existence of a multistable state. The limit cycle shows eddy/swirling motion while the chaotic attractor is a translatory motion, and the initial conditions are crucial in deciding the final state.  \\
\\
The transient for both the initial conditions is $80K$, which is much higher than the $b = 0.18$ case. The transient chaos lasts nearly $80K$ in the case of the third choice of initial condition throughout the range of $cc$. Whereas comparing the Figures(\ref{bfx_1998NL_10}) and (\ref{CH6F13}) shows transient chaos lasts beyond $80K$ for low and intermediate values of $cc$ but dies much earlier for a higher value when the final state is a limit cycle in the case of the second set of initial condition. \\
\\

\begin{figure}[hbt!]
\begin{center}
{\includegraphics[width = 3.5in,height = 2.2in]{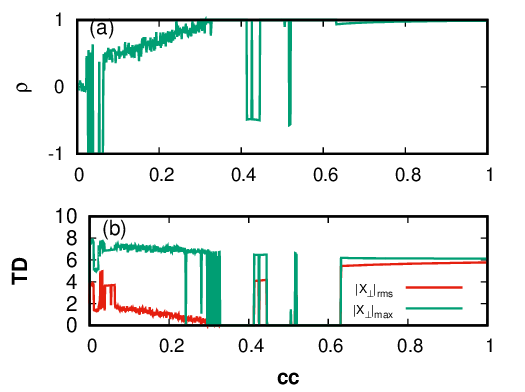}}
\end{center}
\caption{ Pearson coefficient ($\rho$) and Transverse distance (TD) plot for the mutually coupled Thomas system with nonlinear diffusive coupling with $b=0.1998$ and with $10\%$ variation in the initial condition after every cycle of $cc$, Transient time 80k.}
\label{CH6F15}
\end{figure}
\noindent To check the synchronized state, we calculated the synchronization measures up to the transient time $150K$. Then, the final state was a limit cycle state with the nature of synchronization as space lag hence means eddy motion. Even after removing the transient up to $150K$, the asymptotic state remains a limit cycle but with no change in synchronization. Now also, the system shows eddy motion. So, we conclude that the synchronization that we obtain at $80K$ is asymptotic. Like in the previous case,  when the dynamics are limit cycles, the synchronized state is lag instead of CS. This result clearly shows bifurcation from chaos to limit cycle changes nature of synchronization from CS to LS / ALS or space lag.\\
\\
\begin{figure}[hbt!]
\begin{center}
{\includegraphics[width = 3.0in,height = 1.8in]{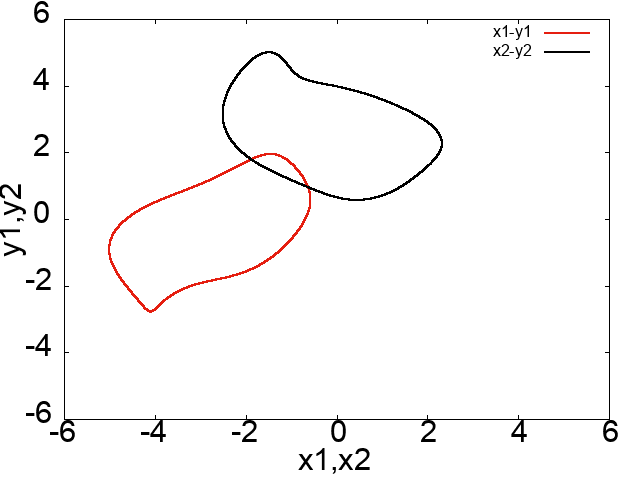}}
\end{center}
\caption{The projected state space of particle one and two at $cc=0.7$ for the mutually coupled Thomas system with nonlinear diffusive coupling with $b=0.1998$ and transient time 150k and second set of the initial conditions. }
\label{CH6F13B}
\end{figure}
\section{\label{sc5}Results and Discussions}
Comparing the results for the two coupling schemes, it is clear that non-linearity in the coupling function adds extra features in the synchronized states. It is known that there is a bifurcation from CS to desynchronization for a higher value of coupling constant in two linearly coupled R\"ossler oscillators\cite{Sum11}. The LTE cuts at two points on the coupling axis. However, this is not the case here for non-linear coupling. In this case, LTE predicts CS throughout, after the onset value of $cc$, and is verified from the result of linear coupling. This prediction is not obeyed in the case of non-linear coupling. The reason for this can be understood based on the stability dynamics of the synchronized manifold given by equations (\ref{stability_linear}) and (\ref{CH6E3}) respectively. Equation(\ref{stability_linear}) is linear whereas  Equation(\ref{CH6E3}) is non-linear. In other words,  Equation(\ref{stability_linear}) is the linearized form of  Equation(\ref{CH6E3}). The LTE is based on the linearization of the  Jacobian matrix. Linearization of  Equation(\ref{CH6E3})  leads to a similar result as that of Equation(\ref{stability_linear}),  which does not capture a complete picture due to the presence of a non-linear $ cosine $ term. Therefore we carried out a full calculation of the transverse Lyapunov exponent instead of finding eigenvalues of the linearized Jacobian. Therefore we can capture the different effects of non-linear stability equations for synchronization.\\
\\
Nonlinearity gives rise to dramatic effects. It produces bifurcation from CS to LS or ALS and CS(in the case of the multistable case) at some intermediate value of $cc$. The LS and ALS were observed earlier only in linearly coupled autonomous non-identical chaotic oscillators, identical autonomous oscillators with delay coupling\cite{gon04}, or linearly coupled identical non-autonomous oscillators\cite{ss1}. In all these cases, LS and ALS appear before the onset of CS only. In the present case, these two forms of synchronizations are observed between two stable CS or after CS. The calculations of similarity function and bifurcation diagram confirm LS and ALS. Similar to the linear case, in this case, also a weak form of synchronization(amplitude envelope) is observed for very weak coupling where systems remain in the chaotic regime for $b = 0.18$. The appearance of a weak form of synchronization before the onset of CS is due to the system's specific characteristics. It is there for both linear and non-linear couplings.
\\
\\
Two mutually coupled Thomas oscillators undergo a period of transient chaos for both linear(weak momentum transfer) and nonlinear(substantial momentum transfer) coupling.
Although there are no transient attractors in uncoupled Thomas oscillators, they appear (mostly transient chaos) when two oscillators are mutually coupled. The transient state's lifetime increases if the coupling's nature changes from linear to sinusoidal nonlinear. It also increases when $b=0.1998$, keeping the same initial condition. The coupled system bifurcates from transient chaos to an asymptotic limit cycle for some window of low coupling in all the cases considered in this study. In this window of low coupling values, the coupled system does not achieve complete synchronization(CS) but shows a weak form of synchronization in some cases. The onset of CS occurred for lower coupling constant in the case of $b = 0.1998$ than $b = 0.18$. Both fall in the intermediate range. CS is achieved during a transient chaotic state and remains so for the final asymptotic chaotic state.\\
\\
In the case of $b = 0.18$, with nonlinear coupling, the transient state is a limit cycle for higher values of coupling constant, slightly beyond the onset of CS. During this transient period, the coupled system is in a “ space-lag synchronized state ”(spatially separated with different orientations and correlated time behaviour). The transient time persists before it bifurcates to a stable chaotic CS state. Unlike the case of $b = 0.18$, the non-linearly coupled system for $b = 0.1998$	also bifurcates from transient chaos to a stable limit cycle for some small window of coupling in the intermediate range but beyond the onset of	CS. In this limit cycle state, oscillators show time-lag synchronization. There is no transient state before the final asymptotic limit cycle for coupling far from the onset of CS.	Dynamics with one set of initial conditions remain chaotic in the CS state for a higher coupling value. However, with different initial conditions, the system settles to an asymptotic limit cycle for the same higher value of coupling and in a space lag synchronized state. This limit cycle attractor occupies different regions in the phase space than the chaotic asymptotic state for the other initial condition. This event shows the existence of multi-stable synchronized states in the case of $b = 0.1998$ with nonlinear coupling.\\
\\
Our result is significant in understanding the collective motion of coupled Brownian particles in a dynamic medium. In such cases, there is a velocity-velocity correlation. When variables in our model represent velocity components, there is indeed a possibility of phase correlation (LS or ALS) among the velocities of Brownian particles other than moving with joint velocity(CS). With the linear coupling, one can observe the directed motion of particles just like molecular diffusion, while the collective movement is complex for the nonlinear interaction. For nonlinear cases, it combines translatory, pedalling, and eddy motion via transient chaos. The particle separation increases with the strength of momentum transfer. So the particle is diffusing much faster in a dynamic environment via transient chaos. Swirling motions are responsible for momentum transport in a turbulent flow. It is a form of diffusion known as turbulent diffusion and causes mixing. This kind of swirling motion is responsible for temperature averaging and energy transfer in the fluid. In many natural and artificial phenomena, particle motion mimics the diffusion of particles in a dynamic fluid that can be analyzed and interpreted due to transient chaos.   Contamination by particulates in hot air and flowing water are two examples of such events and are of significant concern as far as the purity of air and water is concerned. Minor particle contamination is a severe cause of concern in industries related to the manufacturing of semiconductors, storage devices, and processing of wafers for the integrated circuit.
\section{\label{sc6}Conclusion}
This paper considers the dynamics of two mutually coupled Thomas oscillators under linear and non-linear coupling schemes. The synchronization thresholds for both these cases were analyzed. We could produce different results not observed earlier. Like synchronization does not get affected by transient state in all the cases when an asymptotic state is also a chaotic trajectory. When there is a transition from transient chaos to an asymptotic limit cycle, there is a change in the nature of synchronization, as observed in the case of $b = 0.1998$ with non-linear coupling. The appearance of lag/ anti-lag synchronizations and space lag within the regime of complete synchronization and transient limit cycle to chaos are due to non-linear coupling. Chiral phenomenon and structure that are observed in fluids and plasmas remain a puzzle due to its intricate formation. Here at least numerically, we could demonstrate that such a structure can be explained with the Thomas model and harmonic momentum transfer between particles. Studying the synchronization property of two coupled Thomas oscillators is the first step to understanding phenomena in many natural systems. The stochastic dynamics of Brownian motion can also be modelled with the chaotic dynamics of the Thomas system. 
\section*{Acknowledgements}
The authors thanks Prof.J.C.Sprott, Professor, Department of Physics, University of Wisconsin for shedding light on transient chaotic phenomena.
\bibliography{vinesh_paper1}

\end{document}